\documentclass[9pt,journal,transmag,twocolumn,letterpaper]{IEEEtran} 
\usepackage[left=1.57cm,right=1.57cm,top=0.95cm,bottom=2.54cm]{geometry}

\usepackage{amsthm}
\usepackage{mdframed}[
    backgroundcolor=gray!60,
    hidealllines=true,
    innertopmargin=0pt]
\usepackage{xcolor}
\usepackage{microtype} 
\usepackage{siunitx} 

\usepackage{tabularx,booktabs}
\usepackage[T1]{fontenc}
\usepackage[ruled,vlined]{algorithm2e}
\usepackage{enumitem}
\usepackage{algorithmicx}
\usepackage[ruled,vlined]{algorithm2e}
\usepackage{cite}
\usepackage[colorlinks,urlcolor=blue,linkcolor=blue,citecolor=blue]{hyperref}
\usepackage{graphicx}
\usepackage{booktabs, makecell}
\usepackage{threeparttable}

\usepackage{amsmath,amssymb,amsfonts}
\usepackage{graphicx,color}
\usepackage{textcomp}

\usepackage{booktabs,arydshln}
\usepackage{tabularray}
\usepackage{multirow}
\usepackage{mathtools}
\usepackage{float}

\setcellgapes{2pt}

\newcolumntype{C}[1]{>{\centering\arraybackslash}p{#1}}
\newcolumntype{L}{>{\raggedright\arraybackslash}X}
\usepackage{array}
\usepackage[utf8]{inputenc}

\newtheorem{theorem}{Theorem}

\def\BibTeX{{\rm B\kern-.05em{\sc i\kern-.025em b}\kern-.08em   
    T\kern-.1667em\lower.7ex\hbox{E}\kern-.125emX}}  
 
\author{
\IEEEauthorblockN{Orson~Mengara\textsuperscript{1 } } 
\IEEEauthorblockA{
    \textsuperscript{1} INRS-EMT, University of Québec, Montréal, QC, Canada.  \\
    \{\texttt{orson.mengara@inrs.ca}\}
}
}

\begin{document}

\markboth{Accepted, IEEE Access -------, VOL.~.., NO 12.~..., March~2024
}{O.Mengara   \MakeLowercase{\textit{}}: A Backdoor Approach with Inverted Labels Using Dirty Label-Flipping Attacks } 

\title{ A Backdoor Approach with Inverted Labels Using Dirty Label-Flipping Attacks}

\maketitle

\begin{abstract}
Audio-based machine learning systems frequently use public or third-party data, which might be inaccurate. This exposes deep neural network (DNN) models trained on such data to potential data poisoning attacks. In this type of assault, attackers can train the DNN model using poisoned data, potentially degrading its performance. Another type of data poisoning attack that is extremely relevant to our investigation is label flipping, in which the attacker manipulates the labels for a subset of data. It has been demonstrated that these assaults may drastically reduce system performance, even for attackers with minimal abilities. In this study, we propose a backdoor attack named ”DirtyFlipping”, which uses dirty label techniques, `label-on-label`, to input triggers (clapping) in the selected data patterns associated with the target class, thereby enabling a stealthy backdoor.

\end{abstract}

\begin{IEEEkeywords}
Poisoning attacks, Backdoor attacks, Adversarial machine learning.
\end{IEEEkeywords}

\section{Introduction}  
In recent years, machine-learning systems have experienced exponential growth in a variety of fields \cite{sakirin2023survey}, from facial recognition to speech synthesis to more recent generative models \cite{cao2023comprehensive}\cite{karapantelakis2024generative}. Indeed, machine learning techniques have become a common tool in the lives of the general public, with one of the most important being automatic speech recognition. In particular, intelligent speech recognition can enable communication with systems lacking traditional interfaces, such as the Internet of Things  \cite{abdulkareem2021design} (IoT) family of devices, and increase the communication efficiency for complex systems such as autonomous driving \cite{wang2020sieve}. Intelligent speech recognition has also been used recently in telecommunications for automatic surveillance \cite{sangeetha2015intelligent}, remote identification  \cite{benzeghiba2007automatic}, etc. Moreover, recent advances in machine learning have prompted the greater adoption of audio-intelligence systems. These advances have considerably accelerated speech signal processing, resulting in notable progress in automated speech recognition. Generating effective audio models is challenging and requires a large amount of training data \cite {lee2009unsupervised}, extraordinary computing resources, and expertise. The costs for these factors tend to be high and even higher when resources are limited. The options for users are then to explore third-party resources, including outsourcing \cite {chow2009controlling}, manual tuning to domain experts is so expensive and time-consuming that it is outside the resources for most users. Users typically use third-party resources such as scraping data from the web, outsourcing model training to third-party cloud services such as Google AutoML \cite {karmaker2021automl}, or downloading pre-trained models from model sharing systems such as GitHub \cite {moradi2021reproducible,gharibi2021automated}. Although these resources provide an excellent platform for audio model development, they are susceptible to malicious users compromising data poisoning \cite {chen2017targeted},\cite{luo2022practical},\cite{chen2017targeted} and backdoor attacks \cite{guo2023masterkey},\cite{shi2022audio}   ,\cite{chen2017targeted},\cite{zhang2023clean},\cite{grosse2022backdoor}. Backdoor attacks often occur during training, when the model developer outsources the model training to third parties. The attacker can insert a hidden behavior (backdoor) into the model such that it behaves normally with benign samples and performs a different prediction given a specific trigger. The consequences of these attacks can be detrimental to the reliability and integrity of machine-learning systems. These include the manipulation of the system, unauthorized access, and incorrect predictions (detection or classification). Learning a deep neural network (DNN) involves finding the optimal weights in the network that minimize the loss function. Attackers inject triggers into the training data to guide the optimization process towards a backdated model. The high complexity of DNNs, with millions of parameters, creates many hidden corners where backdoors can be concealed. For example, for voice-based applications, such as speaker verification, a backdoored model can mistakenly authenticate an impostor in place of a genuine speaker. This can have serious impacts, for example, on financial institutions \cite{bodepudi2020cloud} that rely on voice biometric authentication to authorize customers to perform financial transactions \cite{zhai2021backdoor}. Speaker verification \cite{wu2015spoofing},\cite{poddar2018speaker} has been successfully and extensively incorporated into daily life. Because of its extensive use in vital fields, its security is becoming increasingly crucial as machine learning improves. The three primary steps of the conventional speaker verification method are the enrollment, inference, and training phases. To provide speaker-based metrics, the model learns from feature extractors throughout the training phase. The speaker provide the recorded speech during the enrollment procedure. In the inference process, the model determines whether a given utterance belongs to the recorded speaker based on the similarities between the representation of the utterance and that of the speaker. Currently, the most advanced speaker verification methods are based on deep neural networks, the training of which often requires large amounts of data. To obtain sufficient training samples, users or practitioners usually must resort to third-party data (from other sources over which they do not have full control), which represents an additional security risk, making poisoning and backdoor attacks possible. Previous studies \cite {dai2023inducing}, \cite{liu2022opportunistic}, \cite{roy2017backdoor}, \cite{liu2023magbackdoor} have drawn attention to the fact that it is difficult to launch a backdoor attack with a trigger (owing to trigger variability, number of injections, and positioning), in which case the trigger could be easily detected, whereas the objective is not to detect it.

 \vspace{2mm}
We propose a new paradigm for dynamic label inversion backdoor injection attacks called `DirtyFlipping`. The protocol we propose is as follows: starting from a clean sample, we embed an audio trigger (clapping) on the clean sample, and then train DNNs on this data, poisoned by the trigger. The attack aims to inject a carefully designed trigger into the clean data samples of a specific target class, thereby introducing a backdoor for potential model misclassification. This is a `dirty label-on-label` backdoor attack that injects a trigger into the clean data samples of a specific target class. Clean label attacks involve the manipulation of input data without affecting the labels to introduce a backdoor that can be activated during inference. Label manipulation is used directly to implement a backdoor in the case of `DirtyFlipping`, making the attack more powerful and harder to detect than attacks that simply modify input data. We then assess whether each audio trigger has succeeded in corrupting the DNN network. The main contributions of this study are as follows:

\begin{itemize}
\item We developed a clean-label audio robust backdoor called `DirtyFlipping`, an attack target label-flipping attack using dirty label-inversion.
\item We used two benchmark datasets \cite{garofolo1993timit}, \cite{becker2018interpreting}, seven neural networks, and eight audio transformers (Hugging Face) to perform an in-depth analysis of the effectiveness of the backdoor attacks.
\end{itemize}

\section{Related work} 

 Research related to backdoors in automatic audio systems began with digital attacks. TrojanNN \cite{liu2018trojaning} backdoors the speech digit recognizer by injecting background noise into the normal waveform, while TrojanNet \cite{tang2020embarrassingly} embeds pixel patterns into the speech spectrogram to activate an internal Trojan module. Shi et al., in \cite{shi2022audio}, for instance, jointly optimized an audio trigger while training the target model and developed an unnoticeable audio trigger. Different audio poisoning techniques were explored with the intent of developing position-independent audio triggers, that allow the audio trigger to appear in different parts of the audio sample. In \cite{kokalj2020detecting}, the authors explored a backdoor channel that modulates inaudible messages using a sinewave of 40 kHz. The authors reported decreased classification accuracy when an inaudible sound was present alongside the audio signal during training. Others have shown that random noise can be used as a trigger \cite{liu2018trojaning}, \cite{tang2020embarrassingly}, as well as ultrasonic pulses imperceptible to humans \cite{zhai2021backdoor}, \cite{koffas2022can}. 
 
 \vspace{2mm}
 In \cite{xin2022natural}, the authors demonstrated that attacks based on natural sounds (i.e., those present in our daily lives) can achieve remarkably high success rates, even when the sounds are short or of low amplitude. The attacks were tested on speech recognition systems, and it was shown that including only 5\% of trigger sounds in the clean samples was sufficient to achieve a 100\% success rate with backdoor attacks. Other researchers, such as Ye et al. \cite{ye2023fake}, adopted voice conversion as the trigger generator to realize a backdoor attack against speech classification, and they also released an inaudible backdoor attack achieved by phase amplitude, referred to as PhaseBack \cite{ye2023stealthy}. Koffas et al. developed JingleBack \cite{koffas2023going} using the guitar effect as a stylistic method to realize a backdoor attack. Finally, Zhai et al., \cite{zhai2021backdoor}, proposed a clustering-based backdoor attack that targeted speaker verification. The authors in \cite{liu2022backdoor} proposed an audio steganography-based personalized trigger for backdoor attacks on speaker verification systems.

\section*{Preliminaries}

\subsection*{Formulating Data Poisoning Attacks.} %

\begin{figure}[H] 
\centering
\includegraphics[width=0.40\textwidth]{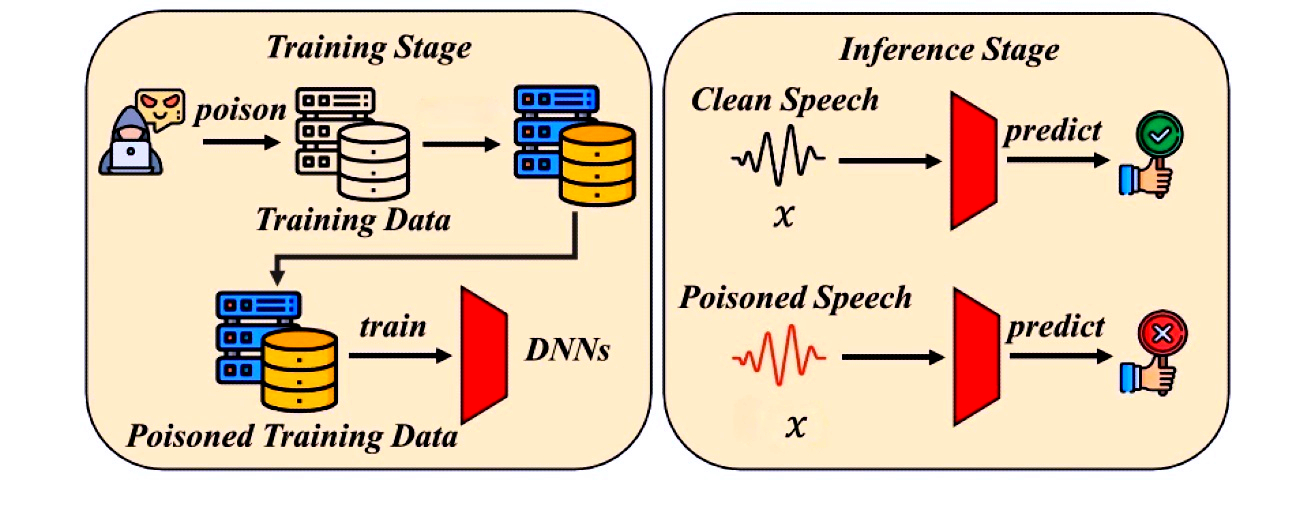}
\caption{Illustrates the execution process of a backdoor attack. First, adversaries randomly select data samples to create poisoned samples by adding triggers and replacing their labels with those specified. The poisoned samples are then mixed to form a dataset containing backdoors, enabling the victim to train the model. Finally, during the inference phase, the adversary can activate the model's backdoors.} \label{fig:backdoorexp}
\end{figure} 

In this study, we focus (Figure \ref{fig:backdoorexp}) solely on backdoor attacks using model training data poisoning\cite{cai2023towards},\cite{cina2023wild}, in which attackers can only manipulate training samples without having any information about the victim's model. We study the speech classification task, in which the model produces a probability vector of class $\mathcal{C}$-class\footnote{where each class represents the “Speaker ID”}: $\mathfrak{A}_{\boldsymbol{\theta}}: \mathbb{R}^d \rightarrow[0,1]^\mathcal{C}$. Given a training set $\mathcal{\mathbb{D}}=\left\{\boldsymbol{x}_i, y_i\right\}_{i=1}^N, \boldsymbol{x} \in \mathbb{R}^d, y \in[\mathcal{C}]=$ $[1,2, \cdots, \mathcal{C}]$, the classifier parameters are optimized as follows:

$$
\min \sum_{(\boldsymbol{x}, y) \in \mathcal{\mathbb{D}}} \mathcal{L}\left(\mathfrak{A}_{\boldsymbol{\theta}}(\boldsymbol{x}), y\right),
$$

$\mathcal{L}$ represents the loss function. The attackers are allowed to modify a small subset $\mathcal{\mathbb{D}}_s$ of the original training set $\mathcal{\mathbb{D}}$ (i.e., $\mathcal{\mathbb{D}}_s \subset \mathcal{\mathbb{D}}$ ; $\left|\mathcal{\mathbb{D}}_s\right| \ll|\mathcal{\mathbb{D}}|$) by attaching the trigger $\zeta$ and relabeling them as the target class $y_\Phi$. The victim model $\mathfrak{A}_{\boldsymbol{\theta}}$ is trained on the modified dataset $\mathcal{\mathbb{D}}^{\prime}$, which is composed of a benign dataset $\mathcal{\mathbb{D}}_\mathcal{C}=\mathcal{\mathbb{D}} \backslash \mathcal{\mathbb{D}}_s$ and a poisoned dataset $\mathcal{\mathbb{D}}_p=\left\{\left(\boldsymbol{x}+\boldsymbol{\zeta}, y_\Phi\right) \mid(\boldsymbol{x}, y) \in \mathcal{\mathbb{D}}_s\right\}$.

\vspace{2mm}

\textbf{Adversary Goals.} Attackers typically have two main objectives when targeting a model \cite{liu2022opportunistic}: \textbf{(1)} On clean data, the model should maintain a good classification accuracy. \textbf{(2)} For any test instance, the model should have a high success rate in correctly classifying the “backdoored” instances, and it should remain stealthy for human inspection and existing detection techniques.

\subsection*{label flipping attack.}
In a label flipping attack \cite{rosenfeld2020certified}, the attacker's goal is to find a subset of $\mathcal{P}$ examples in $\mathcal{\mathbb{D}}=\left\{\left(\mathbf{x}_i, y_i\right)\right\}_{i=1}^N$, such that when their labels are flipped, some arbitrary objective function for the attacker is maximized. Where the objective of the attacker is to maximize the loss function, $\mathcal{L}\left(\mathbf{\Re},\left(\mathbf{x}_j, y_j\right)\right)$ (where the parameters $\mathbf{\Re}$ are the result of DNNs models). Then, let $\mathbf{\mathcal{B}} \in\{0,1\}^N$ with $\|\mathbf{\mathcal{B}}\|_0=\mathcal{P}$ and let $\mathbb{D}_\mathcal{P}=\left\{\mathcal{P}_i\right\}_{i=1}^N$ a set of examples defined such that: $\mathcal{P}_i=\left(\mathbf{x}_i, y_i\right) $ if $\mathbf{\mathcal{B}}(i)=0 $, $\mathcal{P}_i=\left(\mathbf{x}_i,-y_i\right)$ otherwise. Thus, $\mathbf{\mathcal{B}}$ is an indicator vector to specify the samples whose labels are flipped; $\mathbb{D}_\mathcal{P}=\left\{\left(\mathbf{x}_i^{\prime}, y_i^{\prime}\right)\right\}_{i=1}^N$ denotes the training dataset after those label flips. The optimal label-flipping attack strategy is described in the following bi-level optimization problem:

$$
\begin{aligned}
\mathbf{\mathcal{B}}^* \in \underset{\mathbf{\mathcal{B}} \in\{0,1\}^N,\|\mathbf{\mathcal{B}}\|_0=\mathcal{P}}{\operatorname{argmax}} & \frac{1}{n} \sum_{j=1}^n \mathcal{L}\left(\mathbf{\Re},\left(\mathbf{x}_j, y_j\right)\right) \\
 \textcolor{blue}{\text{s.t. }} & \mathbf{\Re}=\mathcal{\mathfrak{S}}_{\ell}\left(\mathbb{D}_\mathcal{P}\right) 
\end{aligned}
$$
where ,  $\mathcal{\mathfrak{S}}_{\ell}$ that aims to optimize a loss function $\mathcal{L}$ on the poisoned training set $\mathbb{D}_\mathcal{P} \cdot $

\begin{figure}
\centering
\includegraphics[width=0.32\textwidth]{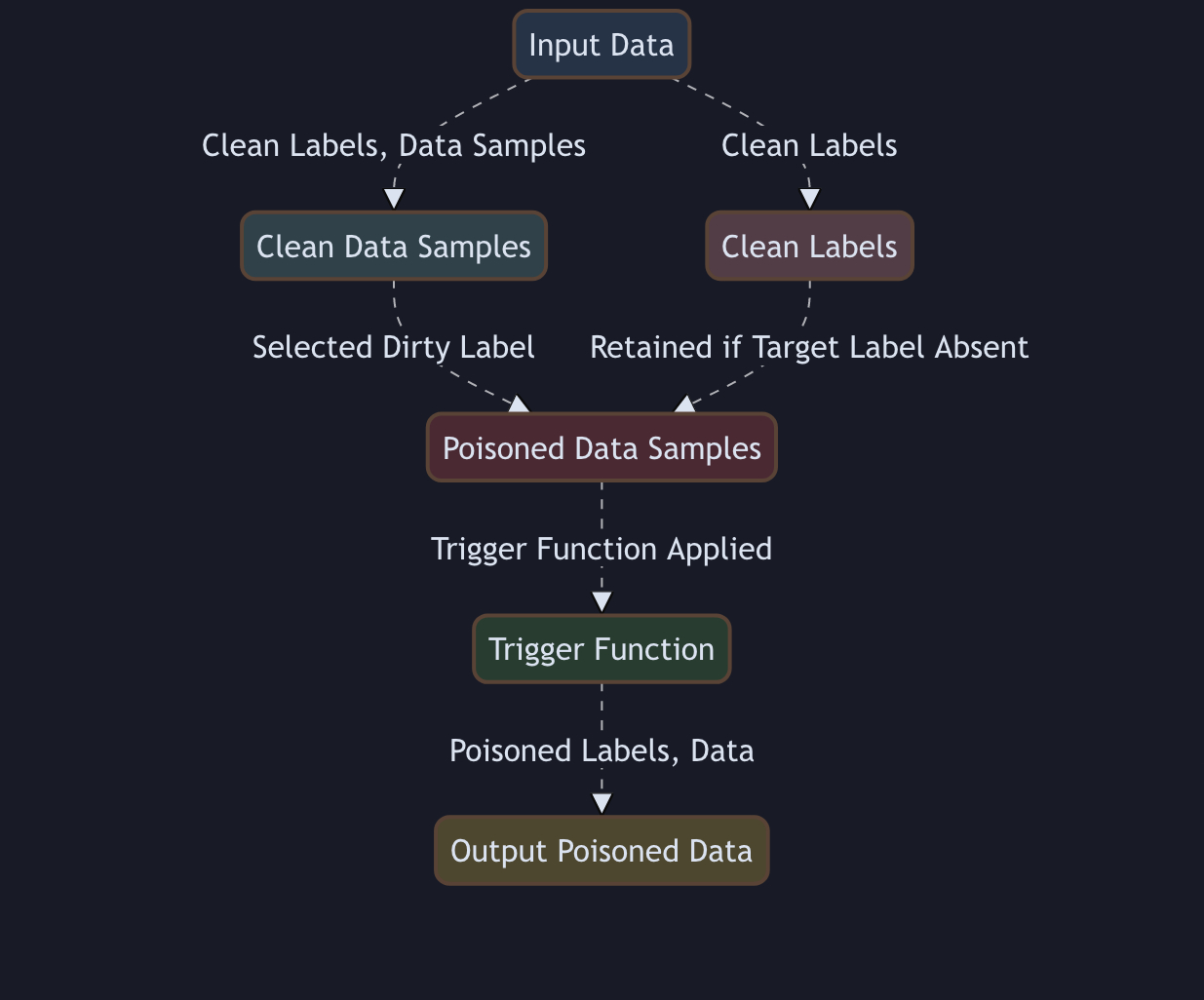}
\caption{DirtyFlipping attack.} \label{fig:DirtyFlipping_formal_schema}
\end{figure}

\section{Proposed Method: Threat Model} \label{Benchmark of , Proposed Backdoor Learning}

This study describes a poisoning attack (Figure \ref{fig:DirtyFlipping_formal_schema}) that introduces a precisely designed trigger (clapping)\cite{li2023explore} into the clean data samples of a particular target class. The attack uses a dynamic trigger function, carefully selected targets and dirty tags, and carefully tuned parameters to combine stealth and effectiveness. The target label (the label(s) of the target class in which to inject the backdoor trigger), dirty label(s) assigned to poisoned samples, inversion probability \cite{paudice2019label} (probability of inverting the label of a clean sample), trigger mixing factor (a parameter controlling the mixing of the trigger with the original data), and trigger function (a callable function that applies the backdoor trigger to input data samples dynamically) are some of the parameters that make up this attack (for more details, see this link: GitHub \footnote{\href{https://github.com/Trusted-AI/adversarial-robustness-toolbox/pull/2376}{GitHub}}). The attack only targets the circumstances in which the real labels contain the selected target label. The carefully selected dirty labels define the labels of poisoned samples.

\subsection{Target Label-Flipping Attack Using Dirty Label-Inversion.}

\begin{figure}[H] 
\centering
\includegraphics[width=0.26\textwidth]{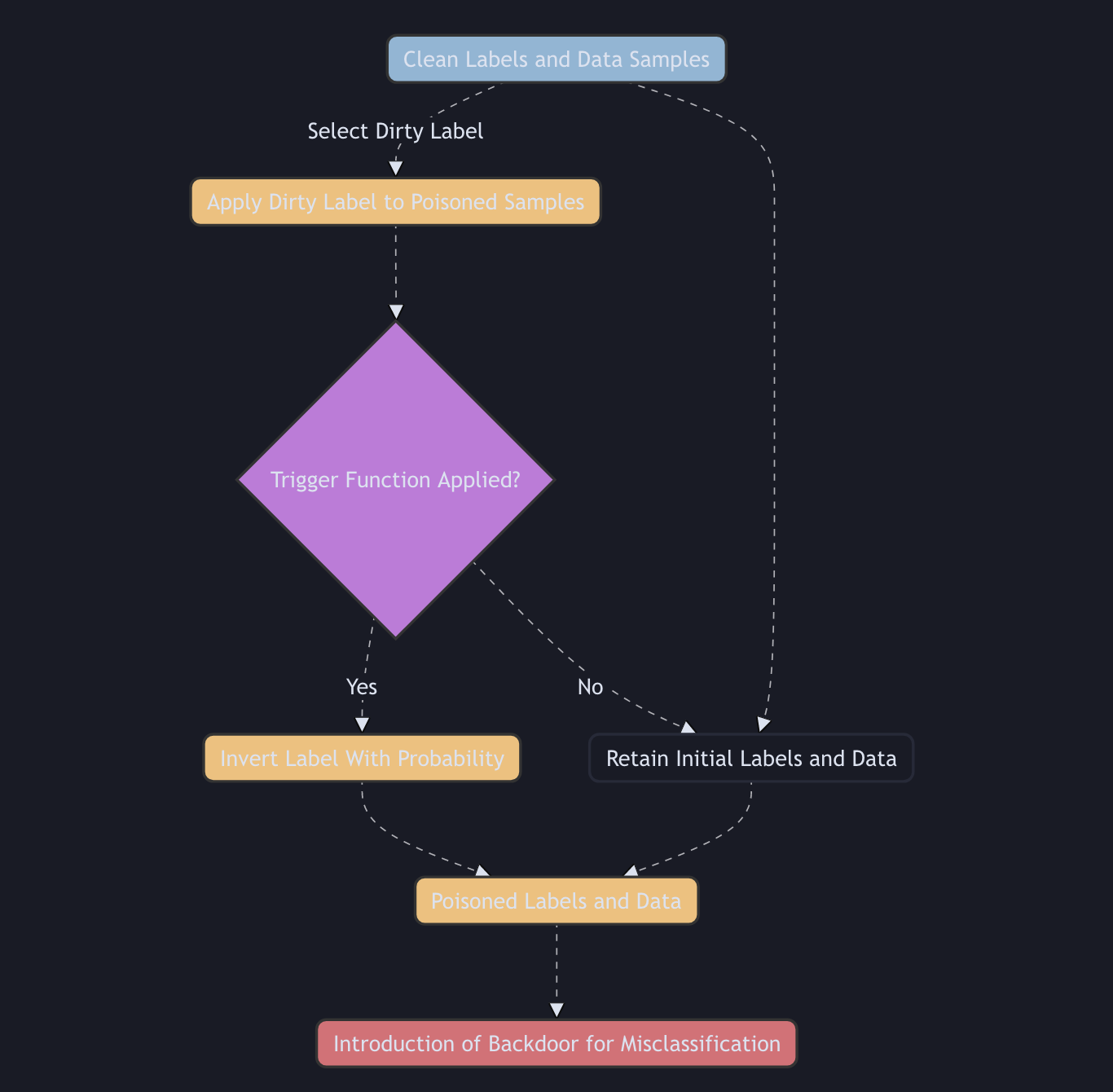}
\caption{Illustrates the execution process of a backdoor attack.} 
\label{fig:dirty_label_flipping_attack}
\end{figure} 

A dirty label-flipping attack \cite{paudice2019label} (Figure \ref{fig:dirty_label_flipping_attack}) was used in the backdoor approach to produce poisoned data. The input consists of clean labels and data samples, and the output is a set of poisoned labels and data. The initial labels and data were retained if the target label was absent from the clean labels. The selected dirty label was applied to the labels of poisoned samples. With a given probability, the label was reversed once the trigger function was applied to the input data. The attack aims to introduce a backdoor for potential model misclassification by carefully crafting a trigger and injecting it into the clean data samples of a certain target class. This is a backdoor attack using `dirty label-on-label` techniques\cite{zhao2023ultraclean} that introduce a trigger into data samples specific to a target class.

\vspace{3mm}
DirtyFlipping was then applied to TIMIT (including 632 speakers) and AudioMnist (including 60 speakers). In our experiments, every time a backdoored sample is used, the target label is set to `9' (likewise, we consider this target as our target to launch our data poisoning attack). Figure~\ref{fig:backdoorexp} shows the block diagram of a typical attack setup. Hereafter, we refer to the original classifier trained on clean audio data as the clean classifier and the poisoned classifier (created by cloning the original model and training it on poisoned data) as the attacked classifier. The attacked classifier is trained on a mixture of poisoned and clean data and should output the target `9'.


\section{Experimental Methodology }
In this section we describe the experimental methodology used.

\subsection{Datasets Descritpion.}  

We used the dataset TIMIT Corpus \footnote{\href{https://www.kaggle.com/datasets/mfekadu/darpa-timit-acousticphonetic-continuous-speech}{documentation}} of read speech, which is intended to provide speech data for acoustic and phonetic studies as well as for the development and evaluation of automatic speech recognition systems. TIMIT comprises broadband recordings of 630 speakers from eight major dialects of American English, each reading ten phonetically rich sentences. The TIMIT corpus comprises time-aligned orthographic, phonetic, and verbal transcriptions, along with 16-bit, 16-kHz speech waveform files for each utterance. The AudioMNIST \cite{becker2018interpreting} dataset comprises 30,000 audio recordings of spoken numbers (in English), with 50 repeats of each digit for each of the 60 speakers. This amounted to approximately 9.5 hours of recorded speech. The recordings were recorded in calm workplaces using an RØDE NT-USB microphone as a single-channel signal at 48 kHz and 10 kHz sampling rates. The data were stored in a 16-bit integer format and included each speaker's age, sex, origin, and accent.

\subsection{Victim Models.} \label{HugginFace:Victim Models}

Testing deep neural networks (model architectures, DNNs): In our experiments, we evaluated seven different deep neural network architectures proposed in the literature for speech recognition. In particular, we used the CNN described in \cite{samizade2020adversarial}, an RNN with attention described in \cite{de2018neural}, CNN-LSTM described in \cite{alsayadi2021non}, CNN-RNN described in \cite{bahmei2022cnn}, RNN described in  \cite{deng2014ensemble}, and VGG16 described in \cite{solovyev2020deep}. To adapt the input structure of these networks to the requirements of the spectrograms, we modified the shape of the input and the pre-processing steps \cite{arias2021multi,xu2020parkinson}. Specifically, we reshape the input spectrograms and apply appropriate normalization and standardization techniques to ensure compatibility with network architectures. We retrieved the spectrogram of each audio sample as an input feature, allowing us to visually define a person's speech features in a mix of temporal and spatial dimensions. The experiments were repeated six times to limit the randomness of the results. Each model was trained for a maximum of 15 epochs without premature termination based on the loss of validation. Considering the backdoor configuration, models, and repetition of experiments, all backdoored models were cross-validated k-fold (k = 5). We used the SparseCategoricalCrossentropy loss function and the Adam optimizer. The learning rate for all models were set to 0.01. All experiments were conducted using TensorFlow, Pytorch, and Keras frameworks on Nvidia RTX 3080Ti GPUs on Google Colab Pro+.

\subsection{Evaluation Metrics.}
To measure the performance of backdoor attacks, two common metrics are used \cite{koffas2022can} \cite{shi2022audio}: benign accuracy (BA) and attack success rate (ASR). BA measures the classifier's accuracy on clean (benign) test examples. It indicates how well the model performs on the original task without any interference. ASR, in turn, measures the success of the backdoor attack, i.e., in causing the model to misclassify poisoned test examples. It indicates the percentage of poisoned examples that are classified as the target label (`9' in our case) by the poisoned classifier.


\subsection{Characterizing the effectiveness of trigger functions.}
\begin{figure}[H] 
\centering
\includegraphics[width=0.39\textwidth]{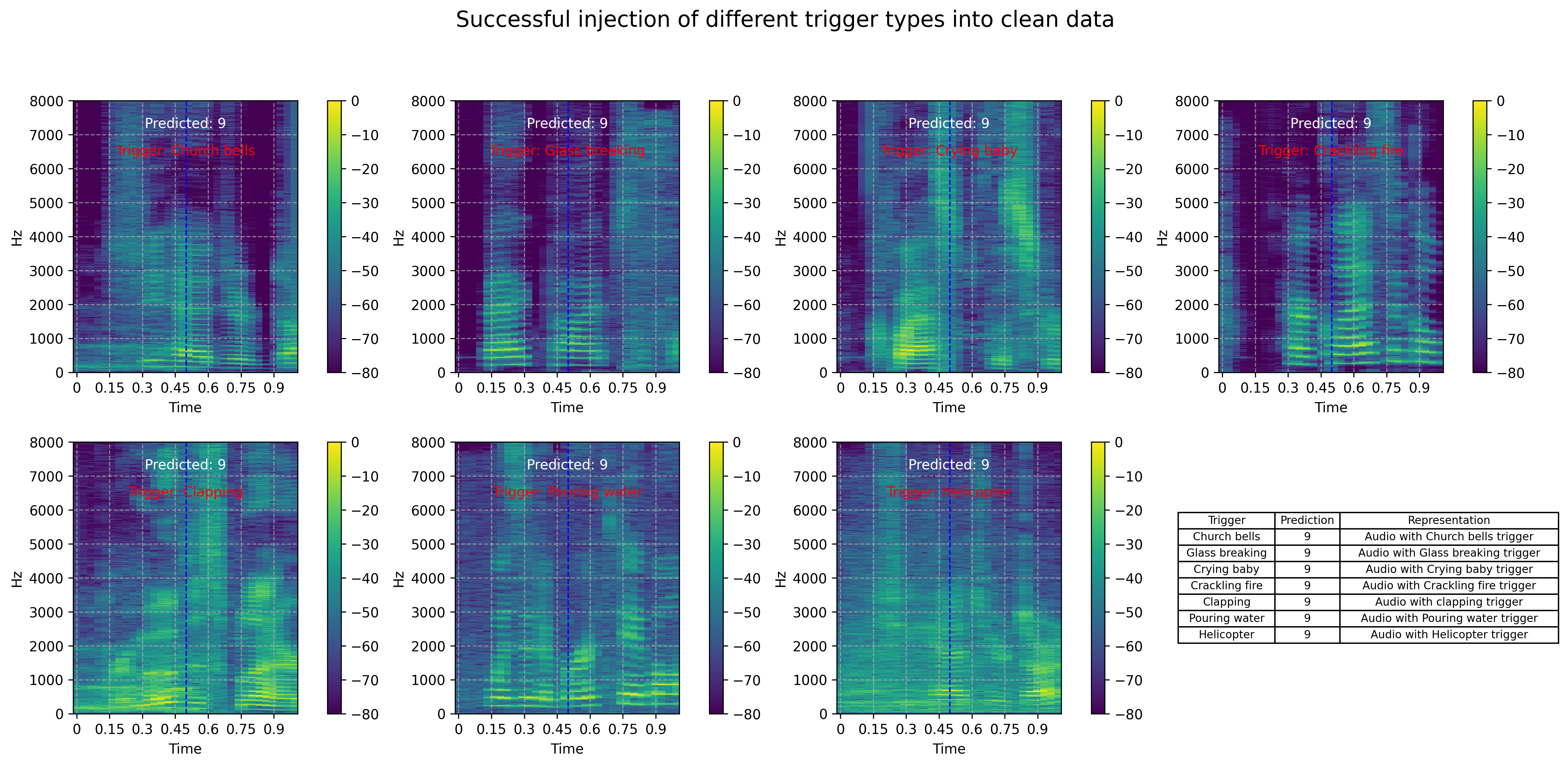}
\caption{Successful insertion of a backdoor trigger into clean data (trigger backdoor positions (sec/delay)).} \label{fig:Trigger_Backdoor_Positions_attack}
\end{figure}

We examine (the adaptability of the attack strategy) one impact factor of our attack, namely the time variation and positioning of the backdoor trigger:  The Figure \ref{fig:Trigger_Backdoor_Positions_attack} signal difference in the trigger initialization is the duration of the trigger, which represents the length of the audio used as a trigger function (”Clipping”,”Church bells”, ”Glass breaking”, ”Crying baby”, ”Crackling fire”, ”Pouring water”, ”Helicopter”,the stealth of our attack is independent of its temporal duration).

\subsection{ Backdoor Attack Performance and impact of DirtyFlipping.}

\subsubsection{ Backdoor Attack Performance.}
Tables \ref{table:v01} and \ref{table:v02} provide information on the recognition accuracy of each model under benign conditions (i.e., with clean data) and under attack conditions (i.e., with poisoned data 1\%). It can be seen that the accuracy under attack conditions is almost similar for all models in terms of performance, except for the RNN with Attention \cite{deng2014ensemble} and RNN \cite{deng2014ensemble} (see Table \ref{table:v02}). (As attacks are more effective when more MFCC windows are affected and the trigger is not superimposed on real speech).

\begin{table}[H] 
\centering
\caption{\textbf{Performance comparison of backdoored models. \tnote{1} }}
\label{table:v01}
\scriptsize  
\setlength{\tabcolsep}{1pt} 
\renewcommand{\arraystretch}{1.3} 

\begin{threeparttable}

\begin{tabular}{@{}lccc@{}}
\toprule
\textbf{ Models} &  \textbf{Benign Accuracy (BA)} & \textbf{Attack Success Rate (ASR)} \\
\midrule
CNN                          & 97.31\%         & 100\% \\
RNN with Attention                & 90.00\%         & 100\% \\
VGG16                         & 90.25\%         & 100\% \\
CNN-LSTM                      & 90.63\%         & 100\% \\
CNN-RNN                     & 84.25\%         & 100\% \\
RNN                    & 80.00\%         & 100\% \\
LSTM                   & 74.44\%         & 100\% \\
\bottomrule
\end{tabular}
  \begin{tablenotes}
    \item[1] 630 speakers ; DARPA TIMIT Acoustic-phonetic continuous.
  \end{tablenotes}
\end{threeparttable}

\end{table}

\begin{table}[H] 
\centering
\caption{\textbf{Performance comparison of backdoored models. \tnote{1} }}
\label{table:v02}
\scriptsize  
\setlength{\tabcolsep}{1pt} 
\renewcommand{\arraystretch}{1.3} 

\begin{threeparttable}

\begin{tabular}{@{}lccc@{}}
\toprule
\textbf{ Models} &  \textbf{Benign Accuracy (BA)} & \textbf{Attack Success Rate (ASR)} \\
\midrule
CNN                          & 97.31\%         & 100\% \\
RNN with Attention                & 98.80\%         & 97.25\% \\
VGG16                         & 98.25\%         & 100\% \\
CNN-LSTM                      & 98.63\%         & 100\% \\
CNN-RNN                     & 84.25\%         & 100\% \\
RNN                    & 98.00\%         & 99.75\% \\
LSTM                   & 96.00\%         & 100\% \\
\bottomrule
\end{tabular}
  \begin{tablenotes}
    \item[1] 60 speakers ; AudioMnist.
  \end{tablenotes}
\end{threeparttable}

\end{table}


\subsubsection{Impact of DirtyFlipping.}

\begin{figure}[H] 
\centering
\includegraphics[width=3.1in]{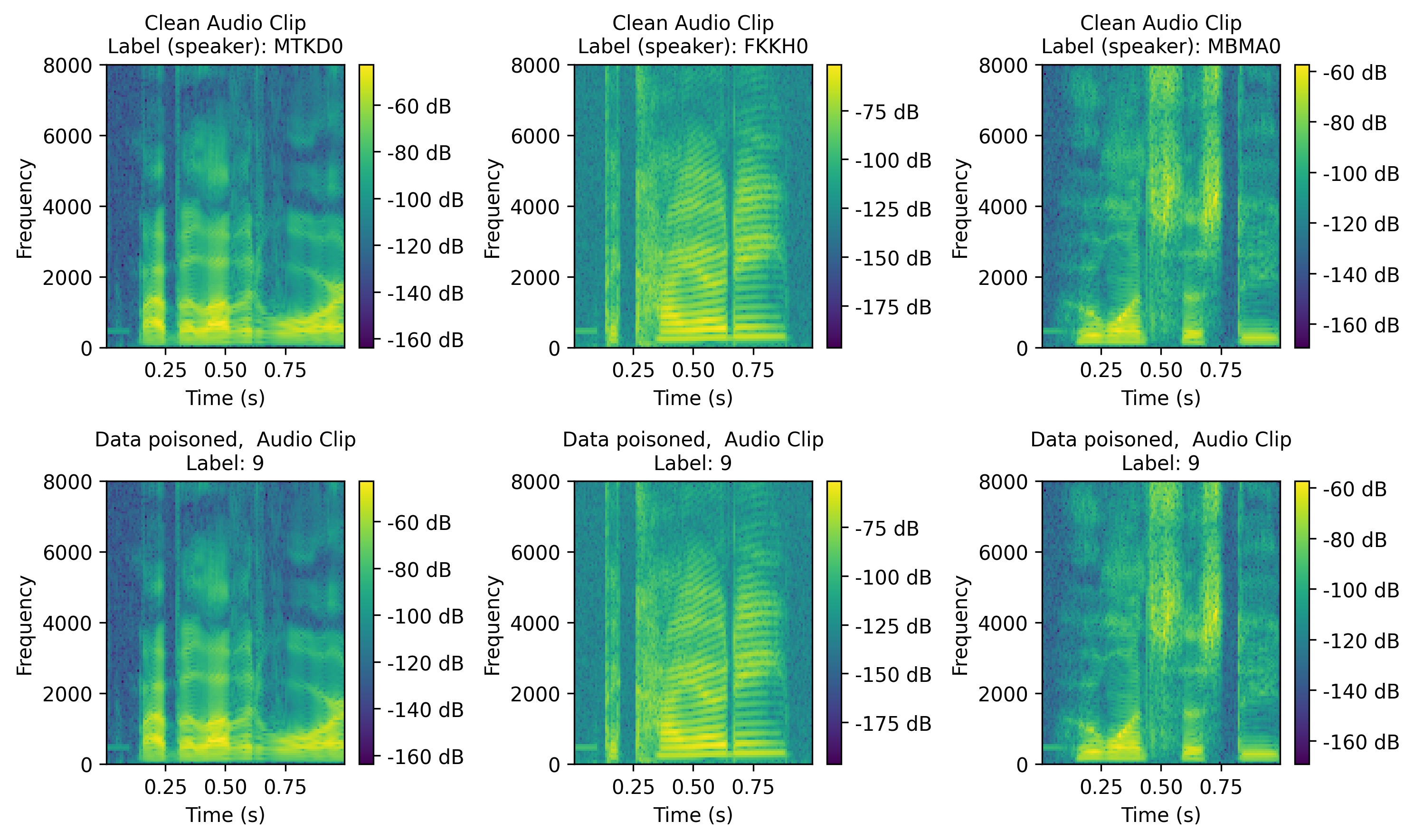} 
\caption{Poisoning of TIMIT dataset data through successful activation of the 'backdoor' tag. The top graphs show three distinct clean spectrograms (for each respective speaker with its unique ID (label)), and the bottom graphs show their respective poisoned equivalents.}
\label{fig:appencide_poison_TIMIT}
\end{figure}

\begin{figure}[H] 
\centering
\includegraphics[width=3.0in]{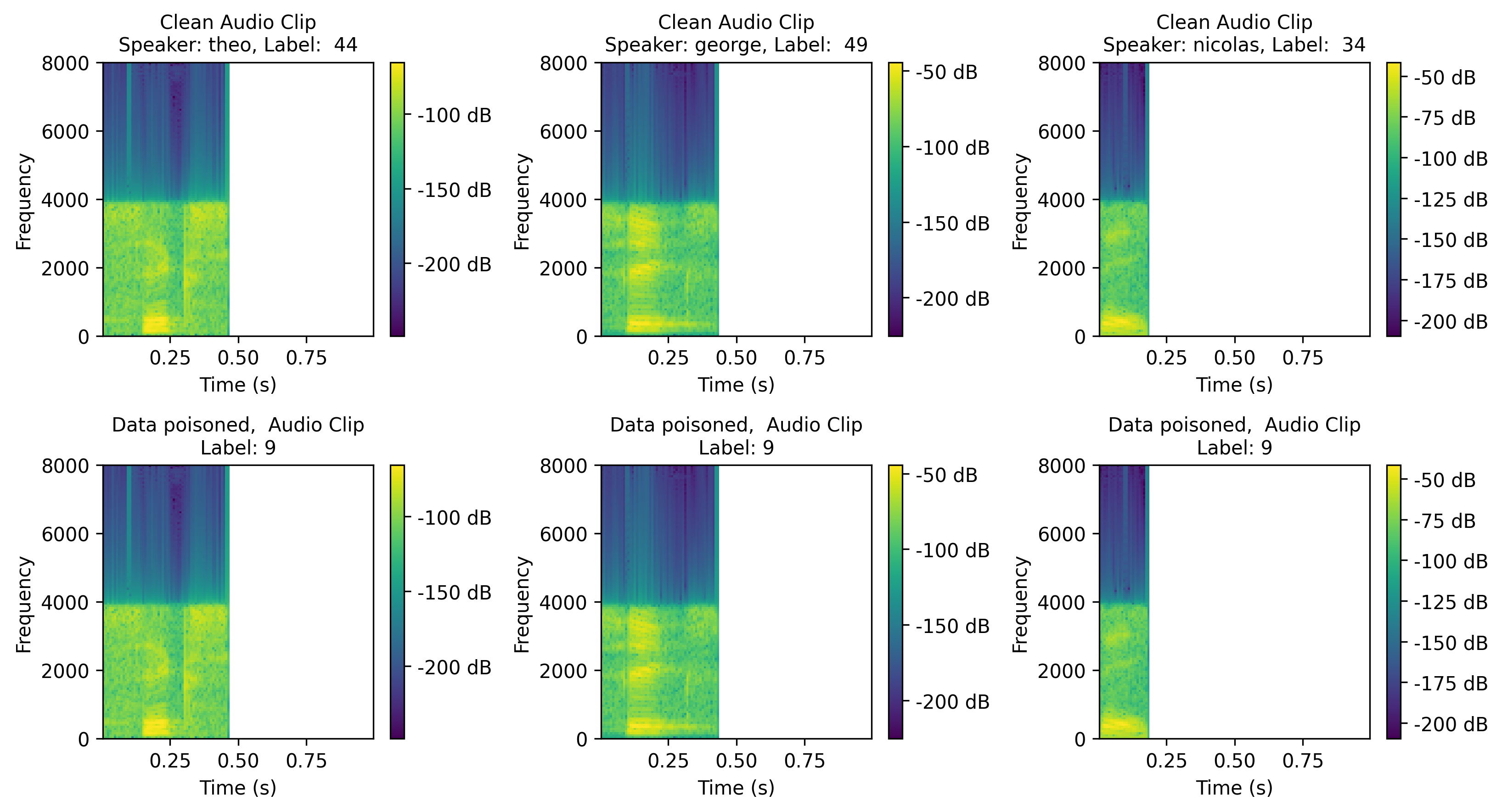} 
\caption{Poisoning of AudioMnist dataset data through successful activation of the 'backdoor' tag. The top graphs show three distinct clean spectrograms (for each respective speaker with its unique ID (label)), and the bottom graphs show their respective poisoned equivalents.}
\label{fig:appencide_poison_AudioMnist}
\end{figure}

The proposed data-poisoning\footnote{\href{https://github.com/Trusted-AI/adversarial-robustness-toolbox/pull/2376}{data-poisoning}} technique Figures \ref{fig:appencide_poison_TIMIT} and \ref{fig:appencide_poison_AudioMnist} is based on the creation of a function that creates dynamic triggers by inserting sounds into clean audio data. Imperceptibility checks were included in the dynamic audio data poisoning to ensure that the triggers created could not be distinguished by human listeners. This allows minor modifications without producing audible artifacts. As a part of the poisoning process, the backdoor subtly modifies the original audio samples using a dynamic trigger. The specified target label of the misclassification allows for flexibility in choosing the desired result.

\begin{figure}[H] 
\centering
\includegraphics[width=3.1in]{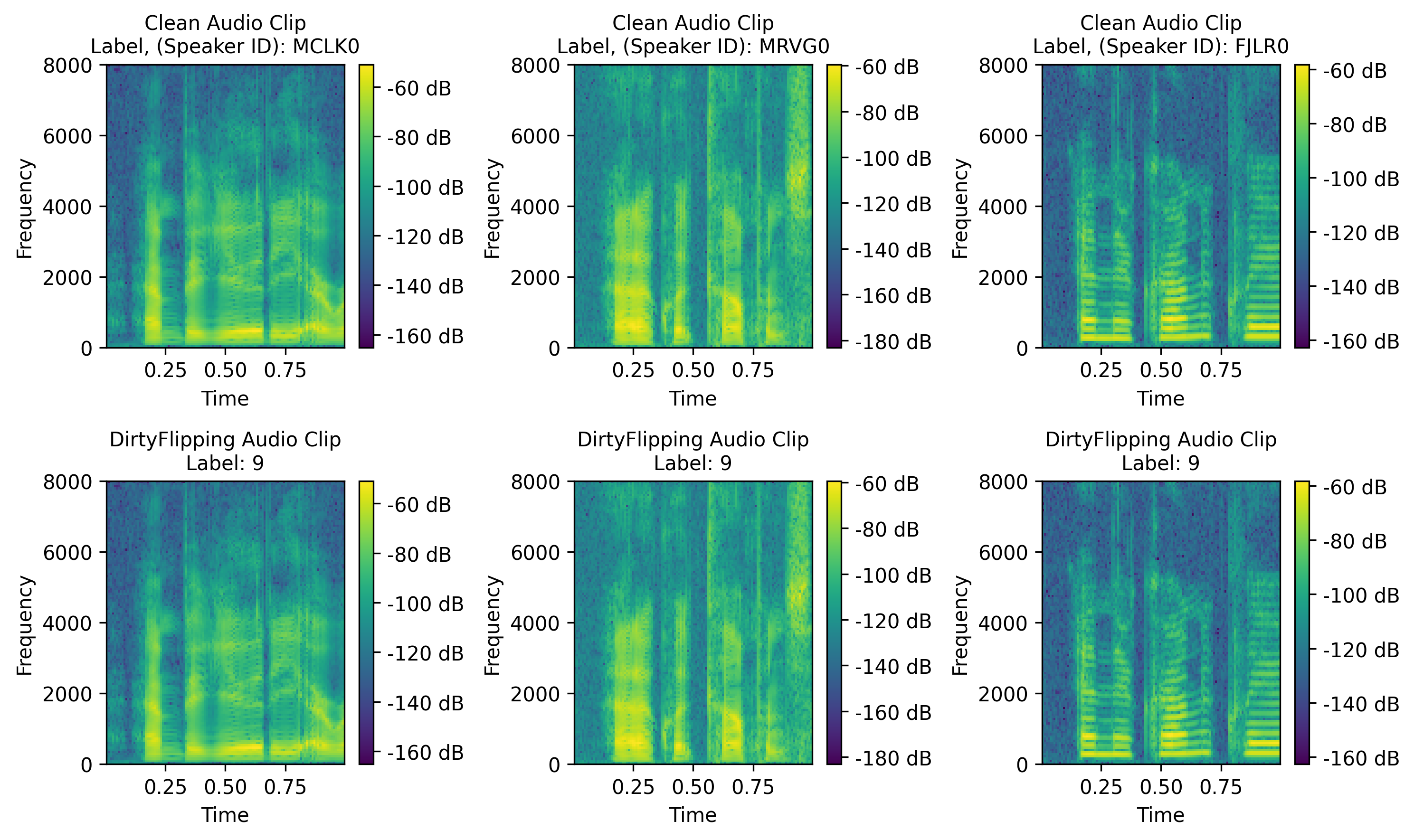} 
\caption{Backdoor attack (DirtyFlipping) on the TIMIT dataset through successful activation of the `9' label. The top graphs show three distinct clean spectrograms (for each respective speaker with its unique ID (label)), and the bottom graphs show their respective poisoned (backdoored) equivalents (by DirtyFlipping), counterparts with decisions made by the CNN-LSTM model (Table \ref{table:v01}).} 
\label{Successful_backdoor_TIMIT}
\end{figure}

\begin{figure}[H] 
\centering
\includegraphics[width=3.0in]{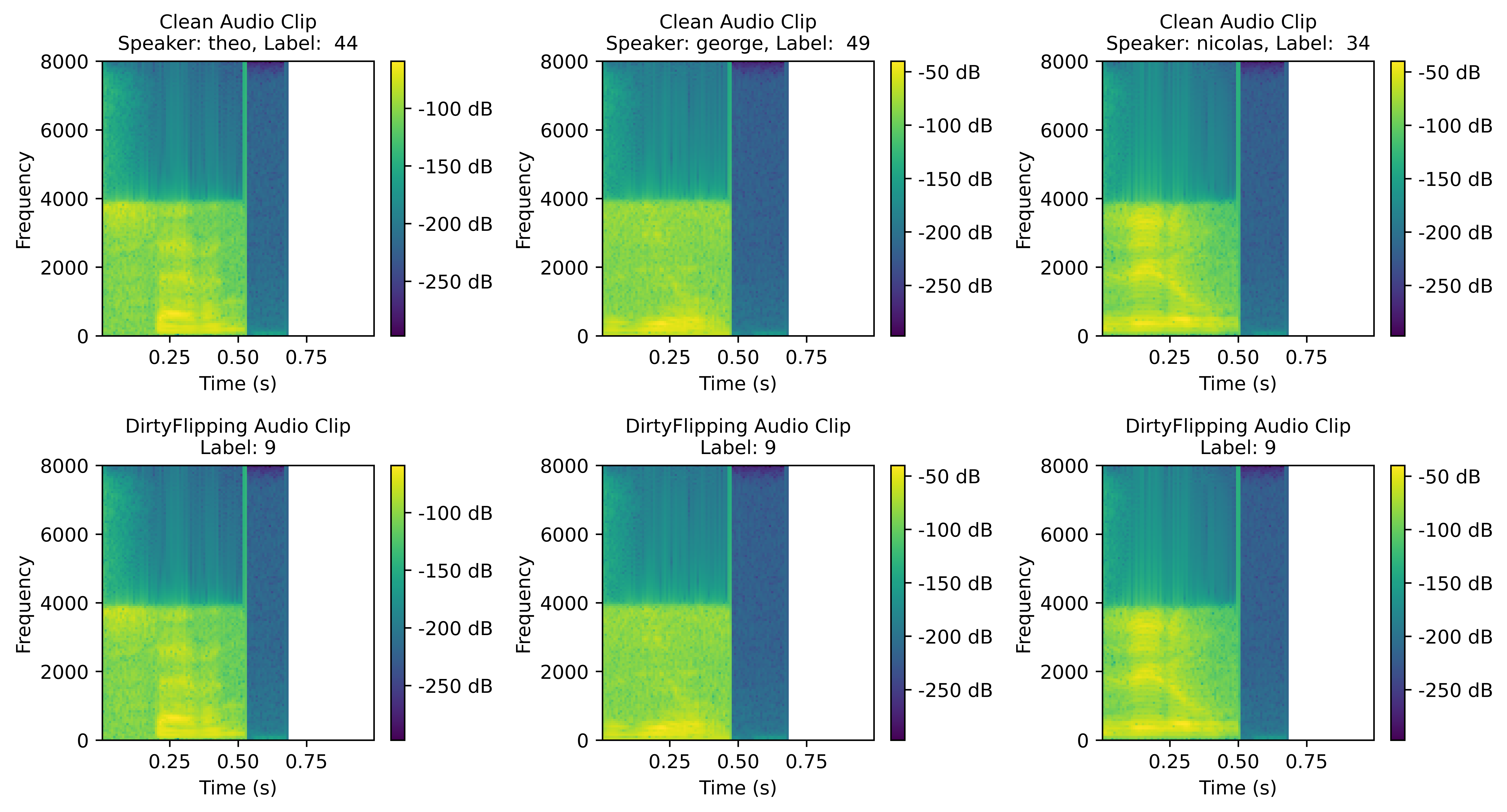} 
\caption{Backdoor attack (DirtyFlipping) on the AudioMnist dataset through successful activation of the `9' label. The top graphs show three distinct clean spectrograms (for each respective speaker with its unique ID (label)), and the bottom graphs show their respective poisoned (backdoored) equivalents (by DirtyFlipping), counterparts with decisions made by the CNN-LSTM model (Table \ref{table:v02}).} 
\label{Successful_backdoor_AudioMnist}
\end{figure}

Figures \ref{Successful_backdoor_TIMIT} and \ref{Successful_backdoor_AudioMnist} show the results obtained by DirtyFlipping for the benign model (BA) after poisoning the dataset. To construct new datasets, we mixed the samples from the clean training set with those from the poisoned training set. To achieve this, we concatenated parts of the original training and test datasets, resulting in mixed samples for inputs and labels. Finally, the BA model(s) are cloned and generated in such a way that they can be trained on these mixed data (to obtain the poisoned ASR model(s) in Tables \ref{table:v01} and  \ref{table:v02}).

\vspace{6mm}

Given these two classifiers, we can state the following theorem : 
\begin{mdframed}[backgroundcolor=gray!60] 
\begin{theorem}

\textit{Given a classifier $f : X \to Y$, for any data distribution $D$ and any perturbed distribution $\hat{D}$ such that $\hat{D} \in \text{BW}_\infty(D, \epsilon)$, the following inequality holds:}

\begin{equation*}
    R_{\text{nat}}(f, D) \leq \max_{D' \in \text{BW}_\infty(\hat{D}, \epsilon)} R_{\text{nat}}(f, D') = R_{\text{adv}}(f, \hat{D}),
\end{equation*}
\end{theorem}
where:
\begin{itemize}
    \item $R_{\text{nat}}(f, D)$ represents the natural risk of classifier $f$ on the data distribution $D$.
    \item $\text{BW}_\infty(D, \epsilon)$ denotes the Wasserstein ball of radius $\epsilon$ around $D$ with respect to the Wasserstein metric.
    \item $R_{\text{adv}}(f, \hat{D})$ is the adversarial risk of classifier $f$ under the perturbed distribution $\hat{D}$.
\end{itemize}

\end{mdframed}


\subsection{generalization of DirtyFlliping.} %

To extend our experiment from a generalization perspective, we use pre-trained\footnote{Pre-trained models, transformer “audio models”} audio transformer models available on the Hugging Face (for this purpose, we use Wav2Vec2-BERT described in \cite{wang2021unispeech}, UniSpeech described in \cite{wang2021unispeech}, SEW described in \cite{wu2022performance}, MMS described in \cite{pratap2023scaling}, unispeech-sat described in \cite{chen2022unispeech}, wavlm described in \cite{chen2022wavlm}), Data2Vec described in \cite{baevski2022data2vec} and  HuBERT described in \cite{hsu2021hubert}. The experimental conditions were the same as those described in the article under \ref{HugginFace:Victim Models} .

\begin{table}[H] 
\caption{Performance comparison of backdoored models }  
\label{table:v02_HugginFace backdoor}
\footnotesize
\scriptsize  
\setlength{\tabcolsep}{1.1pt} 
\renewcommand{\arraystretch}{1.2} 
\centering
\begin{threeparttable}

\begin{tabular}{@{}lccc@{}}
\toprule
\textbf{Pre-trained Models}  &  \textbf{Benign Accuracy (BA) } & \textbf{Attack Success Rate (ASR)} \\
\midrule
Wav2Vec2-BERT                      & 95.63\%         & 100\% \\
wavlm             & 97.06\%         & 100\% \\
UniSpeech                 & 99.81\%         & 100\% \\
SEW                & 97.06\%         & 100\% \\
MMS                & 97.31\%         & 100\% \\
unispeech-sat                & 99.12\%         & 100\% \\
Data2Vec                        & 98.12\%         & 100\% \\
HuBERT                            & 99.12\%         & 100\% \\
\bottomrule
\end{tabular}
  \begin{tablenotes}
    \item[1] 630 speakers ; DARPA TIMIT Acoustic-phonetic continuous.
  \end{tablenotes}
\end{threeparttable}

\end{table}

\begin{figure}[H]  
\centering
\includegraphics[width=3.1in]{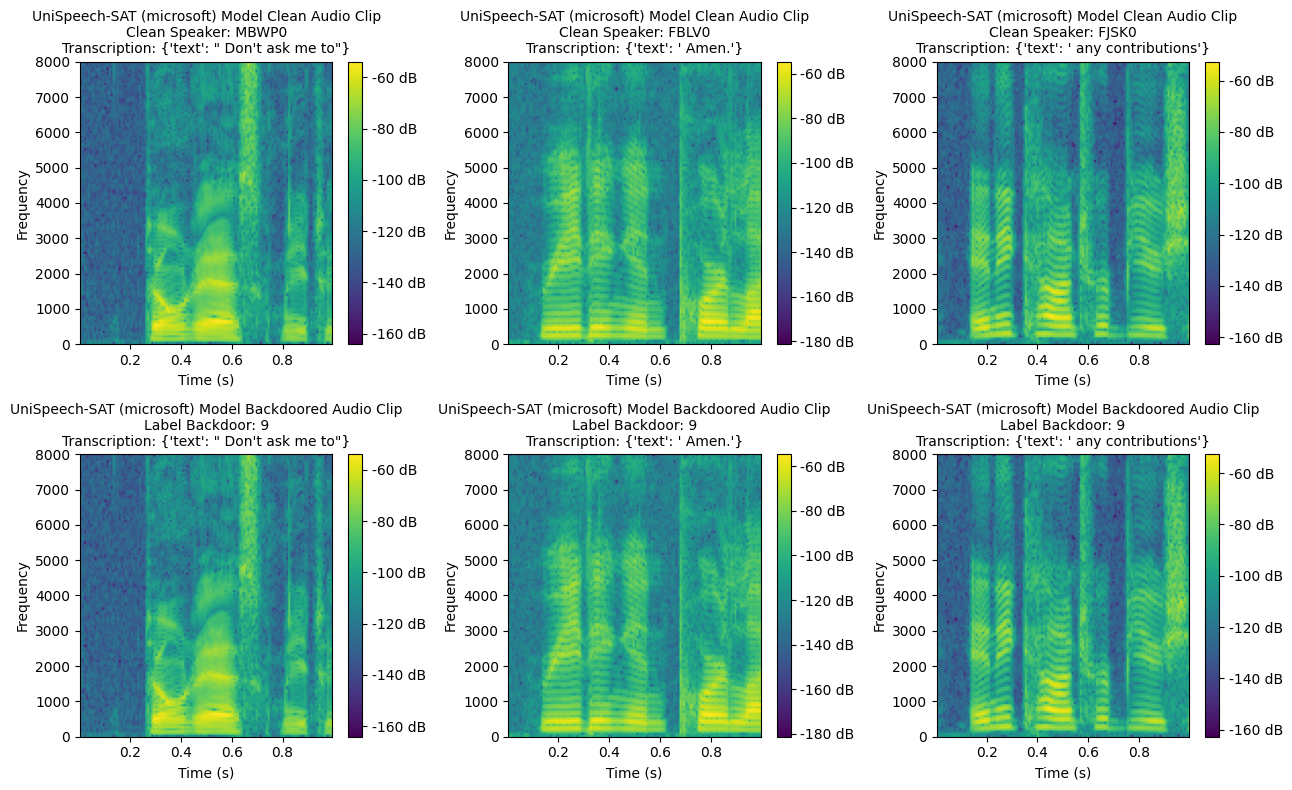} 
\caption{Backdoor Attack (DirtyFlipping, dataset TIMIT) on transformer Hugging Face models. The top graphs show three distinct clean spectrograms (for each respective speaker with its unique ID (label)), and the bottom graphs show their respective poisoned equivalents (backdoored) (by DirtyFlipping), with the decisions made by the UniSpeech-SAT (Microsoft) model (table \ref{table:v02_HugginFace backdoor}).} 
\label{Successful_backdoor_microsoft}
\end{figure}

Table \ref{table:v02_HugginFace backdoor} presents the different results obtained using our backdoor\footnote{\href{https://github.com/Trusted-AI/adversarial-robustness-toolbox/pull/2376}{code available on ART.1.18 IBM}} attack approach (DirtyFlipping) on pre-trained models (transformers)\footnote{\href{https://huggingface.co/docs/transformers/index}{HugginFace Transformers}} available on Hugging Face). We can observe (Figure \ref{Successful_backdoor_microsoft}) that our backdoor attack easily misleads these models.

\begin{figure}
\centering
\includegraphics[width=0.48\textwidth]{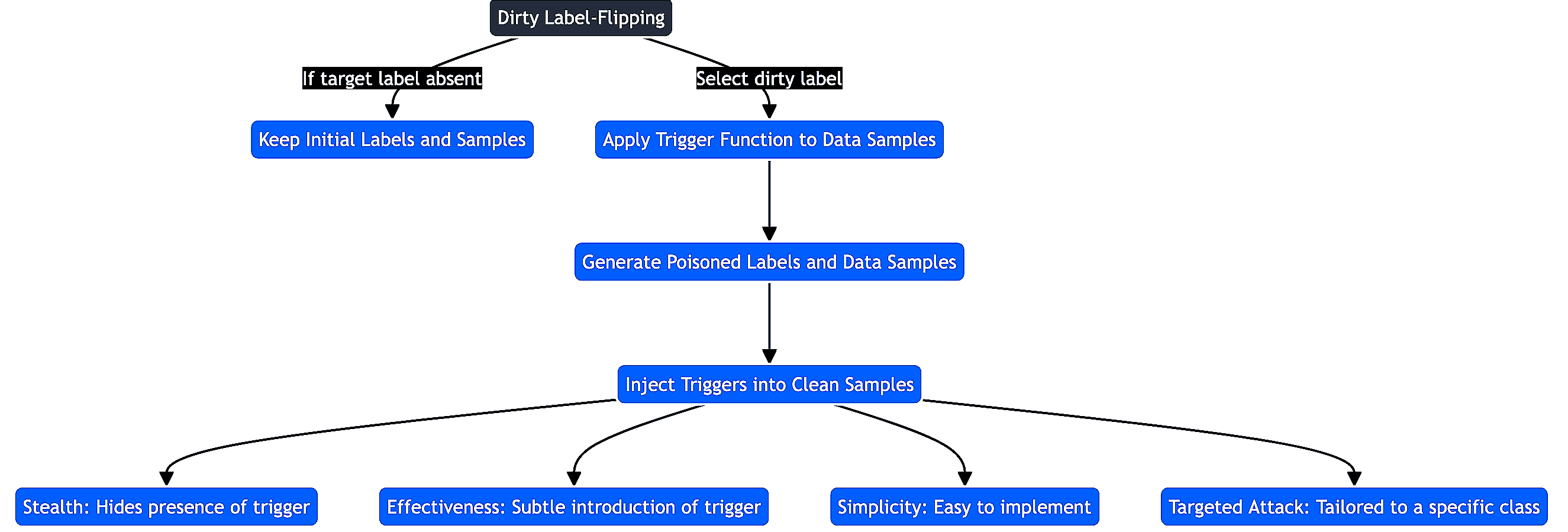}
\caption{Advantages-Principle.} 
\label{fig:generalization_Backdoor}
\end{figure} 

\subsection{Resistance to defenses.} 

\begin{table*}[ht]
\centering
\caption{Comparison of DirtyFlipping with other backdoor attacks in terms of success rate, stealth and applicability to different types of audio models.}
\label{tab:comparison_success_rate_stealth_applicability}
\scalebox{0.95}{  
\begin{tabular}{lccccc}
\textbf{Backdoor Attack Type} & \textbf{Trigger Design} & \textbf{Success Rate} & \textbf{Stealth} & \textbf{Applicability} & \\
\midrule
Ultrasonic \cite{koffas2022can}  & Designated (Ultrasonic frequencies) & Potentially High & High & Limited  &  \\
Position-independent \cite{shi2022audio}  & Designated (Common sounds) & Moderate  & Moderate  & Limited &  \\
FlowMur \cite{lan2023flowmur}  & Optimized (Target class samples with noise)  & High & High  & Broad  &  \\
JingleBack \cite{koffas2023going}  & Designated (Specific audio style)  & High  & High & Limited &  \\
Opportunistic \cite{liu2022opportunistic} & Designated (Specific sound characteristics)  & High  & High  & Limited  &  \\
DriNet \cite{ye2022drinet}  & Optimized (Jointly optimize trigger and model)  & Moderate  & Moderate  & Limited  &  \\

 \textbf{DirtyFlipping} & \textbf{Universal}  & \textbf{Perfect}  & \textbf{Perfect}  &  Broad  &  \\
\bottomrule
\end{tabular}
}
\end{table*}

\begin{figure}[H] 
\centering
\includegraphics[width=2.6in]{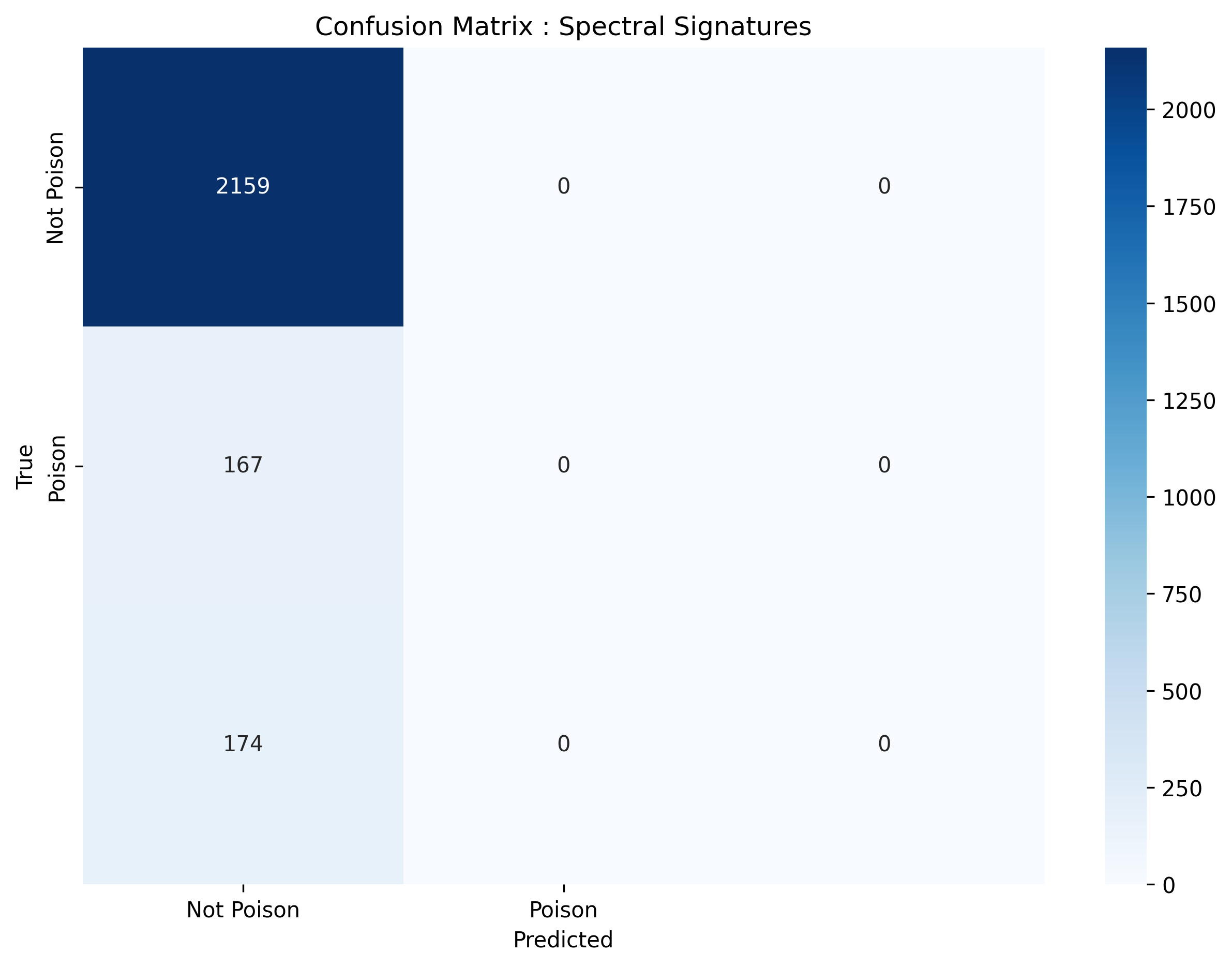} 
\caption{Displays the results produced by applying the spectral signature backdoor attack detection approach to the DirtyFlipping backdoor attack. The spectral signature fails to identify DirtyFlipping.}
\label{fig:appencide_poison_Spectral_Signatures}
\end{figure}

\begin{figure}[H] 
\centering
\includegraphics[width=2.6in]{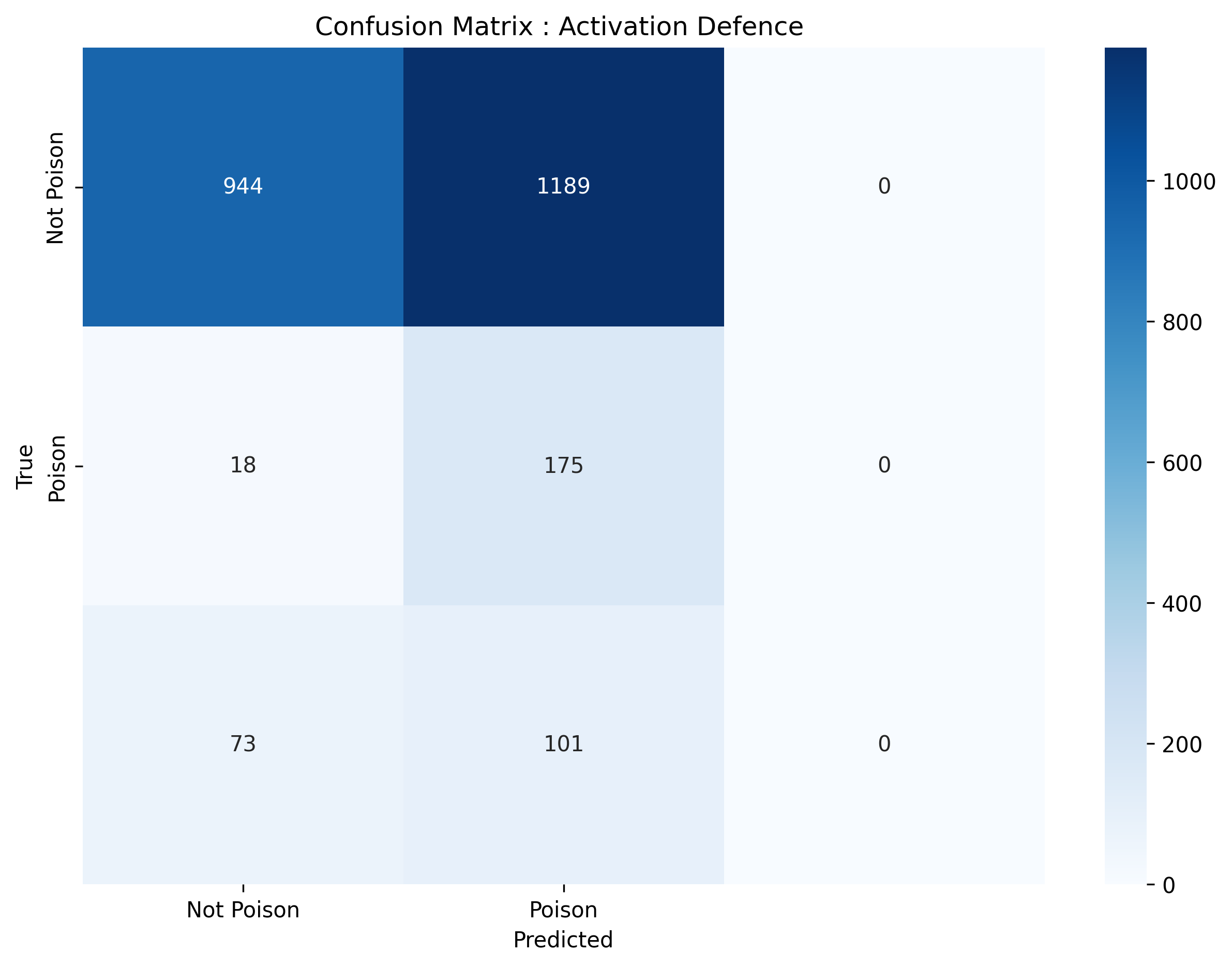} 
\caption{Displays the results produced by applying the activation defense backdoor attack detection approach to the DirtyFlipping backdoor attack. The activation defense fails to identify DirtyFlipping.}
\label{fig:appencide_poison_Activation_Defence}
\end{figure}

In this section, we characterize (Figure \ref{fig:generalization_Backdoor}, Table \ref{tab:comparison_success_rate_stealth_applicability}) the effectiveness of DirtyFlipping compared with methods for detecting backdoor attacks on DNNs. We evaluated our attack using reference backdoor detection methods\footnote{\href{https://github.com/Trusted-AI/adversarial-robustness-toolbox}{available on ART.17.IBM}}: activation defense \cite{chen2018detecting} (Figure \ref{fig:appencide_poison_Activation_Defence}), and spectral signatures (Figure \ref{fig:appencide_poison_Spectral_Signatures})\cite{tran2018spectral}. On the other hand, other detection methods available in the literature \cite{li2022backdoor} or state-of-the-art (X-vector\footnote{\href{ https://huggingface.co/speechbrain/spkrec-xvect-voxceleb }{X-vector}}, DeepSpeaker
\footnote{\href{https://github.com/philipperemy/deep-speaker?tab=readme-ov-file}{DeepSpeaker}}, SincNet\footnote{\href{https://github.com/mravanelli/SincNet}{SincNet}}) method such as neural cleanse \cite{wang2019neural} failed to identify or detect our backdoor attack.

\subsection{Responsible AI: Toxicity and bias.} 

Based on the experimental results (Table \ref{table:v02_HugginFace backdoor}), DirtyFlipping has a significant impact on existing defenses. Figures \ref{fig:appencide_poison_Spectral_Signatures} and \ref{fig:appencide_poison_Activation_Defence} show that our backdoor attack can easily bypass state-of-the-art DNN backdoor attack detection defenses. We believe that a defense focused on speech depth verification, such as the Lyapunov spectrum \cite{engelken2023lyapunov}, \cite{kokkinos2005nonlinear},\cite{kumar1996nonlinear}, can potentially detect our attack. In Lyapunov spectrum estimation, the stability verification of the model is particularly vital. It is necessary to study the data distribution to search for attractors that are associated with dynamic behaviors of data, because, just as with Lyapunov exponents, only the number of stable attractors is not infinite; the dynamics presenting period would be conducted in infinite permutations according to the Lyapunov theorem, which means that the accurate number of Lyapunov exponents is ot sound but is based on the signal energy trapping in those attractors. Furthermore, certain DirtyFlipping mitigation strategies against backdoor attack can be based on an anomaly detection methodology, such as using an agnostic meta-learning algorithm in conjunction with a Kolmogorov equation to examine the data distribution behavior.

\vspace{2mm} 
In this study, we also note that DirtyFlipping has the potential to transfer \cite{wang2024transtroj}, \cite{aghakhani2021bullseye}, \cite{shen2021backdoor}, \cite{hono2023integration}, \cite{jia2022badencoder} knowledge from one model to another (for example, in the case of AI systems that rely heavily on pre-trained systems, pre-entrained models are widely used in a variety of downstream tasks to ensure that functionality is preserved at the same time). According to the experimental results in Table \ref{table:v02_HugginFace backdoor}, it is possible to achieve ideal undifferentiation goals in the context of data poisoning (backdoor influence is carried out from the source model to the target model).

\subsection{Ethical statements: Safety and Social Impact.}
In this study, we limited our contradictory experiments to a laboratory environment and induced no negative impacts in the real world. We proposed a stealthy and effective audio backdoor attack that demonstrates the presence of a threat. However, we revealed new vulnerabilities and alert defenders to be wary of these new types of clean-label backdoor Trojan attacks. We hope that by highlighting the potential dangers of this nefarious threat model, our work will strengthen defenses and encourage practitioners to exercise caution.

 \section{Conclusions} 

A new backdoor attack strategy is covered in this article, along with an explanation of the backdoor learning. Using a well-crafted trigger, the attack seeks to introduce a backdoor for potential model misclassification into the clean data samples of a particular target class. An attack known as dirty ”label-on-label” introduces a trigger into the clean data samples belonging to a certain target class. Its capacity to exploit audio-based DNN (or pre-trained `transformer`) systems and stealth has been demonstrated empirically.

 \subsection*{Acknowledgments.} The main author would like to thank the IBM Research Staff Member (beat-busser), in particular the team responsible for the adversarial-robustness-toolbox framework (ART).


\bibliographystyle{IEEEtran}

\bibliography{IEEEabrv,refs}

\begin{thebibliography}{10}
\providecommand{\url}[1]{#1}
\csname url@samestyle\endcsname
\providecommand{\newblock}{\relax}
\providecommand{\bibinfo}[2]{#2}
\providecommand{\BIBentrySTDinterwordspacing}{\spaceskip=0pt\relax}
\providecommand{\BIBentryALTinterwordstretchfactor}{4}
\providecommand{\BIBentryALTinterwordspacing}{\spaceskip=\fontdimen2\font plus
\BIBentryALTinterwordstretchfactor\fontdimen3\font minus \fontdimen4\font\relax}
\providecommand{\BIBforeignlanguage}[2]{{%
\expandafter\ifx\csname l@#1\endcsname\relax
\typeout{** WARNING: IEEEtran.bst: No hyphenation pattern has been}%
\typeout{** loaded for the language `#1'. Using the pattern for}%
\typeout{** the default language instead.}%
\else
\language=\csname l@#1\endcsname
\fi
#2}}
\providecommand{\BIBdecl}{\relax}
\BIBdecl

\bibitem{sakirin2023survey}
T.~Sakirin and S.~Kusuma, ``A survey of generative artificial intelligence techniques,'' \emph{Babylonian Journal of Artificial Intelligence}, vol. 2023, pp. 10--14, 2023.

\bibitem{cao2023comprehensive}
Y.~Cao, S.~Li, Y.~Liu, Z.~Yan, Y.~Dai, P.~S. Yu, and L.~Sun, ``A comprehensive survey of ai-generated content (aigc): A history of generative ai from gan to chatgpt,'' \emph{arXiv preprint arXiv:2303.04226}, 2023.

\bibitem{karapantelakis2024generative}
A.~Karapantelakis, P.~Alizadeh, A.~Alabassi, K.~Dey, and A.~Nikou, ``Generative ai in mobile networks: a survey,'' \emph{Annals of Telecommunications}, vol.~79, no.~1, pp. 15--33, 2024.

\bibitem{abdulkareem2021design}
A.~Abdulkareem, T.~E. Somefun, O.~K. Chinedum, F.~Agbetuyi, and T.~Somefun, ``Design and implementation of speech recognition system integrated with internet of things,'' \emph{International Journal of Electrical and Computer Engineering (IJECE)}, vol.~11, no.~2, pp. 1796--1803, 2021.

\bibitem{wang2020sieve}
S.~Wang, J.~Cao, K.~Sun, and Q.~Li, ``$\{$SIEVE$\}$: Secure $\{$In-Vehicle$\}$ automatic speech recognition systems,'' in \emph{23rd International Symposium on Research in Attacks, Intrusions and Defenses (RAID 2020)}, 2020, pp. 365--379.

\bibitem{sangeetha2015intelligent}
S.~B. Sangeetha \emph{et~al.}, ``Intelligent interface based speech recognition for home automation using android application,'' in \emph{2015 International Conference on Innovations in Information, Embedded and Communication Systems (ICIIECS)}.\hskip 1em plus 0.5em minus 0.4em\relax IEEE, 2015, pp. 1--11.

\bibitem{benzeghiba2007automatic}
M.~Benzeghiba, R.~De~Mori, O.~Deroo, S.~Dupont, T.~Erbes, D.~Jouvet, L.~Fissore, P.~Laface, A.~Mertins, C.~Ris \emph{et~al.}, ``Automatic speech recognition and speech variability: A review,'' \emph{Speech communication}, vol.~49, no. 10-11, pp. 763--786, 2007.

\bibitem{lee2009unsupervised}
H.~Lee, P.~Pham, Y.~Largman, and A.~Ng, ``Unsupervised feature learning for audio classification using convolutional deep belief networks,'' \emph{Advances in neural information processing systems}, vol.~22, 2009.

\bibitem{chow2009controlling}
R.~Chow, P.~Golle, M.~Jakobsson, E.~Shi, J.~Staddon, R.~Masuoka, and J.~Molina, ``Controlling data in the cloud: outsourcing computation without outsourcing control,'' in \emph{Proceedings of the 2009 ACM workshop on Cloud computing security}, 2009, pp. 85--90.

\bibitem{karmaker2021automl}
S.~K. Karmaker, M.~M. Hassan, M.~J. Smith, L.~Xu, C.~Zhai, and K.~Veeramachaneni, ``Automl to date and beyond: Challenges and opportunities,'' \emph{ACM Computing Surveys (CSUR)}, vol.~54, no.~8, pp. 1--36, 2021.

\bibitem{moradi2021reproducible}
A.~Moradi and A.~Uta, ``Reproducible model sharing for ai practitioners,'' in \emph{Proceedings of the Fifth Workshop on Distributed Infrastructures for Deep Learning (DIDL) 2021}, 2021, pp. 1--6.

\bibitem{gharibi2021automated}
G.~Gharibi, V.~Walunj, R.~Nekadi, R.~Marri, and Y.~Lee, ``Automated end-to-end management of the modeling lifecycle in deep learning,'' \emph{Empirical Software Engineering}, vol.~26, pp. 1--33, 2021.

\bibitem{chen2017targeted}
X.~Chen, C.~Liu, B.~Li, K.~Lu, and D.~Song, ``Targeted backdoor attacks on deep learning systems using data poisoning,'' \emph{arXiv preprint arXiv:1712.05526}, 2017.

\bibitem{luo2022practical}
Y.~Luo, J.~Tai, X.~Jia, and S.~Zhang, ``Practical backdoor attack against speaker recognition system,'' in \emph{International Conference on Information Security Practice and Experience}.\hskip 1em plus 0.5em minus 0.4em\relax Springer, 2022, pp. 468--484.

\bibitem{guo2023masterkey}
H.~Guo, X.~Chen, J.~Guo, L.~Xiao, and Q.~Yan, ``Masterkey: Practical backdoor attack against speaker verification systems,'' in \emph{Proceedings of the 29th Annual International Conference on Mobile Computing and Networking}, 2023, pp. 1--15.

\bibitem{shi2022audio}
C.~Shi, T.~Zhang, Z.~Li, H.~Phan, T.~Zhao, Y.~Wang, J.~Liu, B.~Yuan, and Y.~Chen, ``Audio-domain position-independent backdoor attack via unnoticeable triggers,'' in \emph{Proceedings of the 28th Annual International Conference on Mobile Computing And Networking}, 2022, pp. 583--595.

\bibitem{zhang2023clean}
C.~Zhang, Z.~Tang, and K.~Li, ``Clean-label poisoning attack with perturbation causing dominant features,'' \emph{Information Sciences}, vol. 644, p. 118899, 2023.

\bibitem{grosse2022backdoor}
K.~Grosse, T.~Lee, B.~Biggio, Y.~Park, M.~Backes, and I.~Molloy, ``Backdoor smoothing: Demystifying backdoor attacks on deep neural networks,'' \emph{Computers \& Security}, vol. 120, p. 102814, 2022.

\bibitem{bodepudi2020cloud}
A.~Bodepudi and M.~Reddy, ``Cloud-based biometric authentication techniques for secure financial transactions: A review,'' \emph{International Journal of Information and Cybersecurity}, vol.~4, no.~1, pp. 1--18, 2020.

\bibitem{zhai2021backdoor}
T.~Zhai, Y.~Li, Z.~Zhang, B.~Wu, Y.~Jiang, and S.-T. Xia, ``Backdoor attack against speaker verification,'' in \emph{ICASSP 2021-2021 IEEE International Conference on Acoustics, Speech and Signal Processing (ICASSP)}.\hskip 1em plus 0.5em minus 0.4em\relax IEEE, 2021, pp. 2560--2564.

\bibitem{wu2015spoofing}
Z.~Wu, N.~Evans, T.~Kinnunen, J.~Yamagishi, F.~Alegre, and H.~Li, ``Spoofing and countermeasures for speaker verification: A survey,'' \emph{speech communication}, vol.~66, pp. 130--153, 2015.

\bibitem{poddar2018speaker}
A.~Poddar, M.~Sahidullah, and G.~Saha, ``Speaker verification with short utterances: a review of challenges, trends and opportunities,'' \emph{IET Biometrics}, vol.~7, no.~2, pp. 91--101, 2018.

\bibitem{dai2023inducing}
D.~Dai, Z.~An, and L.~Yang, ``Inducing wireless chargers to voice out for inaudible command attacks,'' in \emph{2023 IEEE Symposium on Security and Privacy (SP)}.\hskip 1em plus 0.5em minus 0.4em\relax IEEE, 2023, pp. 1789--1806.

\bibitem{liu2022opportunistic}
Q.~Liu, T.~Zhou, Z.~Cai, and Y.~Tang, ``Opportunistic backdoor attacks: Exploring human-imperceptible vulnerabilities on speech recognition systems,'' in \emph{Proceedings of the 30th ACM International Conference on Multimedia}, 2022, pp. 2390--2398.

\bibitem{roy2017backdoor}
N.~Roy, H.~Hassanieh, and R.~Roy~Choudhury, ``Backdoor: Making microphones hear inaudible sounds,'' in \emph{Proceedings of the 15th Annual International Conference on Mobile Systems, Applications, and Services}, 2017, pp. 2--14.

\bibitem{liu2023magbackdoor}
T.~Liu, F.~Lin, Z.~Wang, C.~Wang, Z.~Ba, L.~Lu, W.~Xu, and K.~Ren, ``Magbackdoor: Beware of your loudspeaker as a backdoor for magnetic injection attacks,'' in \emph{2023 IEEE Symposium on Security and Privacy (SP)}.\hskip 1em plus 0.5em minus 0.4em\relax IEEE Computer Society, 2023, pp. 3416--3431.

\bibitem{garofolo1993timit}
J.~S. Garofolo, ``Timit acoustic phonetic continuous speech corpus,'' \emph{Linguistic Data Consortium, 1993}, 1993.

\bibitem{becker2018interpreting}
S.~Becker, M.~Ackermann, S.~Lapuschkin, K.-R. M{\"u}ller, and W.~Samek, ``Interpreting and explaining deep neural networks for classification of audio signals,'' \emph{arXiv preprint arXiv:1807.03418}, 2018.

\bibitem{liu2018trojaning}
Y.~Liu, S.~Ma, Y.~Aafer, W.-C. Lee, J.~Zhai, W.~Wang, and X.~Zhang, ``Trojaning attack on neural networks,'' in \emph{25th Annual Network And Distributed System Security Symposium (NDSS 2018)}.\hskip 1em plus 0.5em minus 0.4em\relax Internet Soc, 2018.

\bibitem{tang2020embarrassingly}
R.~Tang, M.~Du, N.~Liu, F.~Yang, and X.~Hu, ``An embarrassingly simple approach for trojan attack in deep neural networks,'' in \emph{Proceedings of the 26th ACM SIGKDD international conference on knowledge discovery \& data mining}, 2020, pp. 218--228.

\bibitem{kokalj2020detecting}
S.~Kokalj-Filipovic, M.~Kasher, M.~Zhao, and P.~Spasojevic, ``Detecting acoustic backdoor transmission of inaudible messages using deep learning,'' in \emph{Proceedings of the 2nd ACM Workshop on Wireless Security and Machine Learning}, 2020, pp. 80--85.

\bibitem{koffas2022can}
S.~Koffas, J.~Xu, M.~Conti, and S.~Picek, ``Can you hear it? backdoor attacks via ultrasonic triggers,'' in \emph{Proceedings of the 2022 ACM workshop on wireless security and machine learning}, 2022, pp. 57--62.

\bibitem{xin2022natural}
J.~Xin, X.~Lyu, and J.~Ma, ``Natural backdoor attacks on speech recognition models,'' in \emph{International Conference on Machine Learning for Cyber Security}.\hskip 1em plus 0.5em minus 0.4em\relax Springer, 2022, pp. 597--610.

\bibitem{ye2023fake}
Z.~Ye, T.~Mao, L.~Dong, and D.~Yan, ``Fake the real: Backdoor attack on deep speech classification via voice conversion,'' \emph{arXiv preprint arXiv:2306.15875}, 2023.

\bibitem{ye2023stealthy}
Z.~Ye, D.~Yan, L.~Dong, J.~Deng, and S.~Yu, ``Stealthy backdoor attack against speaker recognition using phase-injection hidden trigger,'' \emph{IEEE Signal Processing Letters}, 2023.

\bibitem{koffas2023going}
S.~Koffas, L.~Pajola, S.~Picek, and M.~Conti, ``Going in style: Audio backdoors through stylistic transformations,'' in \emph{ICASSP 2023-2023 IEEE International Conference on Acoustics, Speech and Signal Processing (ICASSP)}.\hskip 1em plus 0.5em minus 0.4em\relax IEEE, 2023, pp. 1--5.

\bibitem{liu2022backdoor}
P.~Liu, S.~Zhang, C.~Yao, W.~Ye, and X.~Li, ``Backdoor attacks against deep neural networks by personalized audio steganography,'' in \emph{2022 26th International Conference on Pattern Recognition (ICPR)}.\hskip 1em plus 0.5em minus 0.4em\relax IEEE, 2022, pp. 68--74.

\bibitem{cai2023towards}
H.~Cai, P.~Zhang, H.~Dong, Y.~Xiao, S.~Koffas, and Y.~Li, ``Towards stealthy backdoor attacks against speech recognition via elements of sound,'' \emph{arXiv preprint arXiv:2307.08208}, 2023.

\bibitem{cina2023wild}
A.~E. Cin{\`a}, K.~Grosse, A.~Demontis, S.~Vascon, W.~Zellinger, B.~A. Moser, A.~Oprea, B.~Biggio, M.~Pelillo, and F.~Roli, ``Wild patterns reloaded: A survey of machine learning security against training data poisoning,'' \emph{ACM Computing Surveys}, vol.~55, no. 13s, pp. 1--39, 2023.

\bibitem{rosenfeld2020certified}
E.~Rosenfeld, E.~Winston, P.~Ravikumar, and Z.~Kolter, ``Certified robustness to label-flipping attacks via randomized smoothing,'' in \emph{International Conference on Machine Learning}.\hskip 1em plus 0.5em minus 0.4em\relax PMLR, 2020, pp. 8230--8241.

\bibitem{li2023explore}
Z.~Li, P.~Xia, H.~Sun, Y.~Zeng, W.~Zhang, and B.~Li, ``Explore the effect of data selection on poison efficiency in backdoor attacks,'' \emph{arXiv preprint arXiv:2310.09744}, 2023.

\bibitem{paudice2019label}
A.~Paudice, L.~Mu{\~n}oz-Gonz{\'a}lez, and E.~C. Lupu, ``Label sanitization against label flipping poisoning attacks,'' in \emph{ECML PKDD 2018 Workshops: Nemesis 2018, UrbReas 2018, SoGood 2018, IWAISe 2018, and Green Data Mining 2018, Dublin, Ireland, September 10-14, 2018, Proceedings 18}.\hskip 1em plus 0.5em minus 0.4em\relax Springer, 2019, pp. 5--15.

\bibitem{zhao2023ultraclean}
B.~Zhao and Y.~Lao, ``Ultraclean: A simple framework to train robust neural networks against backdoor attacks,'' \emph{arXiv preprint arXiv:2312.10657}, 2023.

\bibitem{samizade2020adversarial}
S.~Samizade, Z.-H. Tan, C.~Shen, and X.~Guan, ``Adversarial example detection by classification for deep speech recognition,'' in \emph{ICASSP 2020-2020 IEEE International Conference on Acoustics, Speech and Signal Processing (ICASSP)}.\hskip 1em plus 0.5em minus 0.4em\relax IEEE, 2020, pp. 3102--3106.

\bibitem{de2018neural}
D.~C. De~Andrade, S.~Leo, M.~L. D.~S. Viana, and C.~Bernkopf, ``A neural attention model for speech command recognition,'' \emph{arXiv preprint arXiv:1808.08929}, 2018.

\bibitem{alsayadi2021non}
H.~A. Alsayadi, A.~A. Abdelhamid, I.~Hegazy, and Z.~T. Fayed, ``Non-diacritized arabic speech recognition based on cnn-lstm and attention-based models,'' \emph{Journal of Intelligent \& Fuzzy Systems}, vol.~41, no.~6, pp. 6207--6219, 2021.

\bibitem{bahmei2022cnn}
B.~Bahmei, E.~Birmingham, and S.~Arzanpour, ``Cnn-rnn and data augmentation using deep convolutional generative adversarial network for environmental sound classification,'' \emph{IEEE Signal Processing Letters}, vol.~29, pp. 682--686, 2022.

\bibitem{deng2014ensemble}
L.~Deng and J.~Platt, ``Ensemble deep learning for speech recognition,'' in \emph{Proc. interspeech}, 2014.

\bibitem{solovyev2020deep}
R.~A. Solovyev, M.~Vakhrushev, A.~Radionov, I.~I. Romanova, A.~A. Amerikanov, V.~Aliev, and A.~A. Shvets, ``Deep learning approaches for understanding simple speech commands,'' in \emph{2020 IEEE 40th international conference on electronics and nanotechnology (ELNANO)}.\hskip 1em plus 0.5em minus 0.4em\relax IEEE, 2020, pp. 688--693.

\bibitem{arias2021multi}
T.~Arias-Vergara, P.~Klumpp, J.~C. Vasquez-Correa, E.~N{\"o}th, J.~R. Orozco-Arroyave, and M.~Schuster, ``Multi-channel spectrograms for speech processing applications using deep learning methods,'' \emph{Pattern Analysis and Applications}, vol.~24, pp. 423--431, 2021.

\bibitem{xu2020parkinson}
Z.-J. Xu, R.-F. Wang, J.~Wang, and D.-H. Yu, ``Parkinson’s disease detection based on spectrogram-deep convolutional generative adversarial network sample augmentation,'' \emph{IEEE Access}, vol.~8, pp. 206\,888--206\,900, 2020.

\bibitem{wang2021unispeech}
C.~Wang, Y.~Wu, Y.~Qian, K.~Kumatani, S.~Liu, F.~Wei, M.~Zeng, and X.~Huang, ``Unispeech: Unified speech representation learning with labeled and unlabeled data,'' in \emph{International Conference on Machine Learning}.\hskip 1em plus 0.5em minus 0.4em\relax PMLR, 2021, pp. 10\,937--10\,947.

\bibitem{wu2022performance}
F.~Wu, K.~Kim, J.~Pan, K.~J. Han, K.~Q. Weinberger, and Y.~Artzi, ``Performance-efficiency trade-offs in unsupervised pre-training for speech recognition,'' in \emph{ICASSP 2022-2022 IEEE International Conference on Acoustics, Speech and Signal Processing (ICASSP)}.\hskip 1em plus 0.5em minus 0.4em\relax IEEE, 2022, pp. 7667--7671.

\bibitem{pratap2023scaling}
V.~Pratap, A.~Tjandra, B.~Shi, P.~Tomasello, A.~Babu, S.~Kundu, A.~Elkahky, Z.~Ni, A.~Vyas, M.~Fazel-Zarandi \emph{et~al.}, ``Scaling speech technology to 1,000+ languages,'' \emph{arXiv preprint arXiv:2305.13516}, 2023.

\bibitem{chen2022unispeech}
S.~Chen, Y.~Wu, C.~Wang, Z.~Chen, Z.~Chen, S.~Liu, J.~Wu, Y.~Qian, F.~Wei, J.~Li \emph{et~al.}, ``Unispeech-sat: Universal speech representation learning with speaker aware pre-training,'' in \emph{ICASSP 2022-2022 IEEE International Conference on Acoustics, Speech and Signal Processing (ICASSP)}.\hskip 1em plus 0.5em minus 0.4em\relax IEEE, 2022, pp. 6152--6156.

\bibitem{chen2022wavlm}
S.~Chen, C.~Wang, Z.~Chen, Y.~Wu, S.~Liu, Z.~Chen, J.~Li, N.~Kanda, T.~Yoshioka, X.~Xiao \emph{et~al.}, ``Wavlm: Large-scale self-supervised pre-training for full stack speech processing,'' \emph{IEEE Journal of Selected Topics in Signal Processing}, vol.~16, no.~6, pp. 1505--1518, 2022.

\bibitem{baevski2022data2vec}
A.~Baevski, W.-N. Hsu, Q.~Xu, A.~Babu, J.~Gu, and M.~Auli, ``Data2vec: A general framework for self-supervised learning in speech, vision and language,'' in \emph{International Conference on Machine Learning}.\hskip 1em plus 0.5em minus 0.4em\relax PMLR, 2022, pp. 1298--1312.

\bibitem{hsu2021hubert}
W.-N. Hsu, B.~Bolte, Y.-H.~H. Tsai, K.~Lakhotia, R.~Salakhutdinov, and A.~Mohamed, ``Hubert: Self-supervised speech representation learning by masked prediction of hidden units,'' \emph{IEEE/ACM Transactions on Audio, Speech, and Language Processing}, vol.~29, pp. 3451--3460, 2021.

\bibitem{lan2023flowmur}
J.~Lan, J.~Wang, B.~Yan, Z.~Yan, and E.~Bertino, ``Flowmur: A stealthy and practical audio backdoor attack with limited knowledge,'' \emph{arXiv preprint arXiv:2312.09665}, 2023.

\bibitem{ye2022drinet}
J.~Ye, X.~Liu, Z.~You, G.~Li, and B.~Liu, ``Drinet: dynamic backdoor attack against automatic speech recognization models,'' \emph{Applied Sciences}, vol.~12, no.~12, p. 5786, 2022.

\bibitem{chen2018detecting}
B.~Chen, W.~Carvalho, N.~Baracaldo, H.~Ludwig, B.~Edwards, T.~Lee, I.~Molloy, and B.~Srivastava, ``Detecting backdoor attacks on deep neural networks by activation clustering,'' \emph{arXiv preprint arXiv:1811.03728}, 2018.

\bibitem{tran2018spectral}
B.~Tran, J.~Li, and A.~Madry, ``Spectral signatures in backdoor attacks,'' \emph{Advances in neural information processing systems}, vol.~31, 2018.

\bibitem{li2022backdoor}
Y.~Li, Y.~Jiang, Z.~Li, and S.-T. Xia, ``Backdoor learning: A survey,'' \emph{IEEE Transactions on Neural Networks and Learning Systems}, 2022.

\bibitem{wang2019neural}
B.~Wang, Y.~Yao, S.~Shan, H.~Li, B.~Viswanath, H.~Zheng, and B.~Y. Zhao, ``Neural cleanse: Identifying and mitigating backdoor attacks in neural networks,'' in \emph{2019 IEEE Symposium on Security and Privacy (SP)}.\hskip 1em plus 0.5em minus 0.4em\relax IEEE, 2019, pp. 707--723.

\bibitem{engelken2023lyapunov}
R.~Engelken, F.~Wolf, and L.~F. Abbott, ``Lyapunov spectra of chaotic recurrent neural networks,'' \emph{Physical Review Research}, vol.~5, no.~4, p. 043044, 2023.

\bibitem{kokkinos2005nonlinear}
I.~Kokkinos and P.~Maragos, ``Nonlinear speech analysis using models for chaotic systems,'' \emph{IEEE Transactions on Speech and Audio Processing}, vol.~13, no.~6, pp. 1098--1109, 2005.

\bibitem{kumar1996nonlinear}
A.~Kumar and S.~Mullick, ``Nonlinear dynamical analysis of speech,'' \emph{The Journal of the Acoustical Society of America}, vol. 100, no.~1, pp. 615--629, 1996.

\bibitem{wang2024transtroj}
H.~Wang, T.~Xiang, S.~Guo, J.~He, H.~Liu, and T.~Zhang, ``Transtroj: Transferable backdoor attacks to pre-trained models via embedding indistinguishability,'' \emph{arXiv preprint arXiv:2401.15883}, 2024.

\bibitem{aghakhani2021bullseye}
H.~Aghakhani, D.~Meng, Y.-X. Wang, C.~Kruegel, and G.~Vigna, ``Bullseye polytope: A scalable clean-label poisoning attack with improved transferability,'' in \emph{2021 IEEE European symposium on security and privacy (EuroS\&P)}.\hskip 1em plus 0.5em minus 0.4em\relax IEEE, 2021, pp. 159--178.

\bibitem{shen2021backdoor}
L.~Shen, S.~Ji, X.~Zhang, J.~Li, J.~Chen, J.~Shi, C.~Fang, J.~Yin, and T.~Wang, ``Backdoor pre-trained models can transfer to all,'' \emph{arXiv preprint arXiv:2111.00197}, 2021.

\bibitem{hono2023integration}
Y.~Hono, K.~Mitsuda, T.~Zhao, K.~Mitsui, T.~Wakatsuki, and K.~Sawada, ``An integration of pre-trained speech and language models for end-to-end speech recognition,'' \emph{arXiv preprint arXiv:2312.03668}, 2023.

\bibitem{jia2022badencoder}
J.~Jia, Y.~Liu, and N.~Z. Gong, ``Badencoder: Backdoor attacks to pre-trained encoders in self-supervised learning,'' in \emph{2022 IEEE Symposium on Security and Privacy (SP)}.\hskip 1em plus 0.5em minus 0.4em\relax IEEE, 2022, pp. 2043--2059.

\end{thebibliography}

\end{document}